\documentclass[epsf,usegraphicx]{mn2e}

\usepackage{xspace}

\title[The outburst cycle of KL Dra]
{Multi-wavelength observations of the helium dwarf nova KL Dra through 
its outburst cycle}

\author[]
{Gavin Ramsay$^{1}$,  Iwona Kotko$^{2}$, Thomas Barclay$^{1,3}$, C. M.
Copperwheat$^{5}$, \and  Simon Rosen$^{4}$, C. Simon Jeffery$^{1}$,  
T. R. Marsh$^{5}$, Danny Steeghs$^{5}$, Peter J. Wheatley$^{5}$\\
$^{1}$Armagh Observatory, College Hill, Armagh, BT61 9DG\\
$^{2}$Astronomical Observatory, Jagiellonian University, Cracow, Poland\\
$^{3}$Mullard Space Science Laboratory, University College London,
Holmbury St. Mary, Dorking, Surrey, RH5 6NT\\
$^{4}$Department of Physics and Astronomy, University of Leicester, 
University Road, Leicester LE1 7RH\\
$^{5}$Department of Physics, University of Warwick, Coventry, CV4 7AL\\
}

\date{Accepted 2010 May 13.  Received 2010 May 13; in original form
2010 March 15}

\begin{document}
\outer\def\gtae {$\buildrel {\lower3pt\hbox{$>$}} \over 
{\lower2pt\hbox{$\sim$}} $}
\outer\def\ltae {$\buildrel {\lower3pt\hbox{$<$}} \over 
{\lower2pt\hbox{$\sim$}} $}
\newcommand{\ergscm} {ergs s$^{-1}$ cm$^{-2}$}
\newcommand{\ergss} {ergs s$^{-1}$}
\newcommand{\ergsd} {ergs s$^{-1}$ $d^{2}_{100}$}
\newcommand{\pcmsq} {cm$^{-2}$}
\newcommand{\ros} {\sl ROSAT}
\newcommand{\chan} {\sl Chandra}
\newcommand{\xmm} {\sl XMM-Newton}
\newcommand{\swift} {\sl Swift}
\def\rchi{{${\chi}_{\nu}^{2}$}}
\newcommand{\Msun} {$M_{\odot}$}
\newcommand{\Mwd} {$M_{wd}$}
\def\Mdot{\hbox{$\dot M$}}
\def\mdot{\hbox{$\dot m$}}
\newcommand{\teff}{\ensuremath{T_{\mathrm{eff}}}\xspace}

\maketitle

\begin{abstract}

We present multi-wavelength observations of the helium-dominated
accreting binary KL Dra which has an orbital period of 25 mins.  Our
ground-based optical monitoring programme using the Liverpool
Telescope has revealed KL Dra to show frequent outbursts. Although our
coverage is not uniform, our observations are consistent with the
outbursts recurring on a timescale of $\sim$60 days. Observations made
using {\swift} show that the outbursts occur with a similar amplitude
at both UV and optical energies and a duration of 2 weeks. 
Although KL Dra is a weak X-ray
source we find no significant evidence that the X-ray flux varies over
the course of an outburst cycle. We can reproduce the main features of
the 60 day outburst cycle using the Disc Instability Model and a
helium-dominated accretion flow. Although the outbursts of KL Dra are very 
similar to those of the hydrogen accreting dwarf novae, 
we cannot exclude that they are the AM CVn equivalent of WZ 
Sge type outbursts. With
outbursts occurring every $\sim$2 months, KL Dra is an excellent
target to study helium-dominated accretion flows in general.

\end{abstract}


\section{Introduction}

AM CVn systems are accreting binaries consisting of a white dwarf
primary and a degenerate or semi-degenerate secondary star.  Their
binary orbital periods are extremely short ($<$ 70 mins) and are
almost entirely hydrogen-deficient (see Nelemans 2005 for a recent
review).  As such they are the best sources in which to understand
hydrogen-deficient accretion flows.

There are currently $\sim$26 known (or candidate) AM CVn systems (see
Rau et al. 2010 for the most recent discoveries) of which around half a
dozen have shown optical outbursts with amplitudes of 3--4 mag. These
outbursts are assumed to be similar to those observed in the
hydrogen-dominated accreting dwarf novae. Although the physics of
helium-dominated accretion is poorly understood, outbursts are thought
to be due to instabilities in the accretion disc (eg Smak 1983,
Tsugawa \& Osaki 1997 and Kotko, Lasota \& Dubus 2010).

Since many of the more recently discovered AM CVn systems are
relatively faint ($V\sim$19--20) the outburst characteristics of AM
CVn systems as a whole are not well understood. To better characterise
these properties we have started a monitoring programme using the
Liverpool Telescope to obtain images of the northern AM CVn systems
once a week over the course of at least one year.

\begin{figure*}
\begin{center}
\setlength{\unitlength}{1cm}
\begin{picture}(16,8.5)
\put(0.5,-0.5){\includegraphics{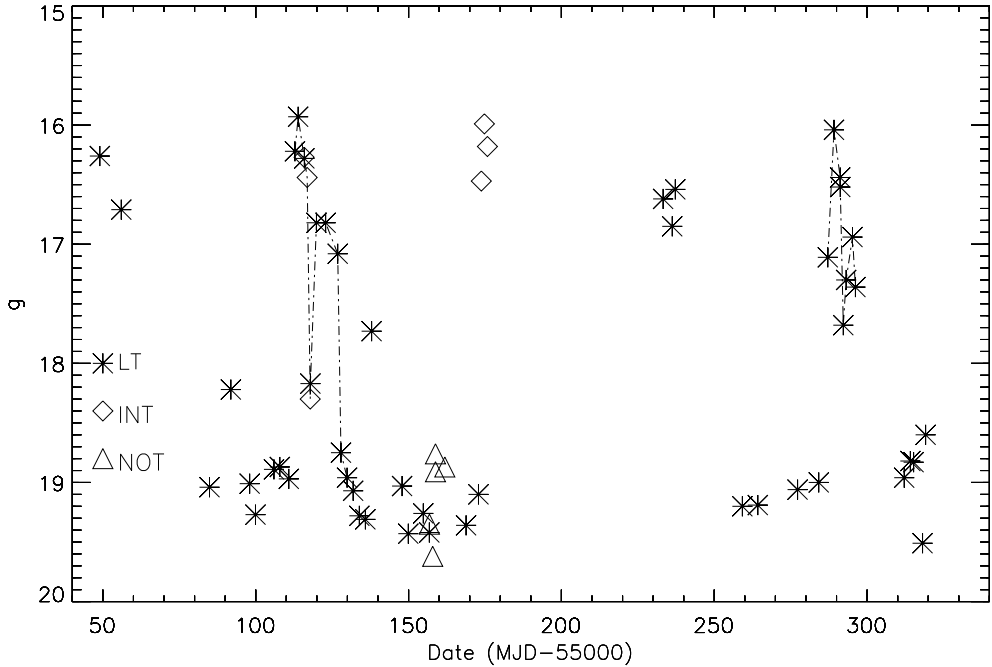}}
\end{picture}
\end{center}
\caption{The optical observations of KL Dra made using the 
Liverpool Telescope (LT), Isaac Newton Telescope (INT) and 
Nordic Optical Telescope (NOT). The first observation was taken on 
2009 July 28th. The dot-dashed line helps guide the eye in the two 
best sampled outbursts.}
\label{lt-summary}
\end{figure*}

The results of our monitoring programme will be presented in a future
paper (Barclay et al. in prep). Here, we concentrate on one individual
system, KL Dra, which was initially thought to be a supernova (Jha et
al 1998). Subsequent observations showed it was an AM CVn system with
an orbital period close to 25 mins (Wood et al. 2002). Our initial
observations made using the Liverpool Telescope showed two outbursts
from KL Dra within 2 months (Barclay, Ramsay \& Steeghs 2009). This
makes KL Dra an excellent system with which to gain a better
understanding of outbursts in AM CVn binaries.

In this paper we present results from our recent observing campaign of
KL Dra using the {\sl Swift} satellite, the Liverpool Telescope, the
Isaac Newton Telescope, the Nordic Optical Telescope, the William
Herschel Telescope and the Gemini North Telescope.

\section{Optical Photometric Observations}
\label{phot}

The strategy for our observations made using the 2.0m Liverpool
Telescope (LT) was to obtain a single 180 sec image of KL Dra using
the RATCAM imager (Steele et al. 2004) in the $g$ band filter
approximately once every week. During outbursts we increased the
sampling rate to once every few days.  Images which had been bias
subtracted and flat-fielded using the LT automatic pipeline were
typically downloaded the afternoon after the observation had been
made. We also obtained supplementary images using the Isaac Newton
Telescope (INT) and the Nordic Optical Telescope (NOT).

Since KL Dra is $5.7^{''}$ from the nucleus of an anonymous galaxy
(Jha et al. 1998 and Fig 2 of Wood et al. 2002), differential imaging
would have been an option for obtaining the photometric light curve of
KL Dra. However, since the images were taken using a number of
telescopes (giving different image scales etc), we decided to use
aperture photometry. Care was taken to exclude as much of the galaxy
as possible and to ensure a star/galaxy free background. The
difference in magnitude between two comparison stars was constant to
within a few 0.01 mag. To place our photometry onto the standard
system, we obtained an image of the field and several standard stars
in Oct 2009 using the INT. This allowed us to determine the $g$ band
magnitude of several local comparison stars.

We show in Figure \ref{lt-summary} our optical light curve of KL Dra.
Our first observation made on 2009 July 28 showed KL Dra in a bright
optical state ($g\sim$16.2). One week later the system had faded by
around $\sim$0.5 mag, after which no observations were obtained for 5
weeks. However, 63 days after our first observation, KL Dra was again
observed in a bright optical state ($g\sim$16.0) with an outburst
amplitude of $\sim$3 mag. The rise to peak brightness (maximum) was
short ($<$2 days) with a short duration drop in brightness $\sim$5
days after maximum. Two weeks after maximum there was another sharp
drop in brightness giving an overall duration for the outburst of
$\sim$15 days.

A third burst was detected from KL Dra 61.5$\pm$2.0 days after the
preceding one. Since KL Dra shortly came too close to the Sun to be
observable, we obtained no ground-based
optical observations for nearly 9 weeks. However, an outburst was
again seen at a time which is consistent with KL Dra showing outbursts
on a repeating interval of $\sim$60 days. Very recently we 
detected the fifth outburst of KL Dra (2010 Apr 1st) which took place
at least 54 days after the start of the previous outburst. Although we 
find no evidence for other periods in our optical data a more comprehensive 
dataset is needed to test this more thoroughly.

During its low optical state, KL Dra appears rather variable which is
probably associated with its orbital period (Wood et al. 2002). It also
shows what appears to be short duration brightenings (up to $\sim$1.5
mags) on several occasions (ie MJD$\sim$55092).

\section{Optical Spectra}
\label{opt-spec}

\begin{figure*}
\begin{center}
\setlength{\unitlength}{1cm}
\begin{picture}(16,10.7)
\put(0,-0.3){\includegraphics{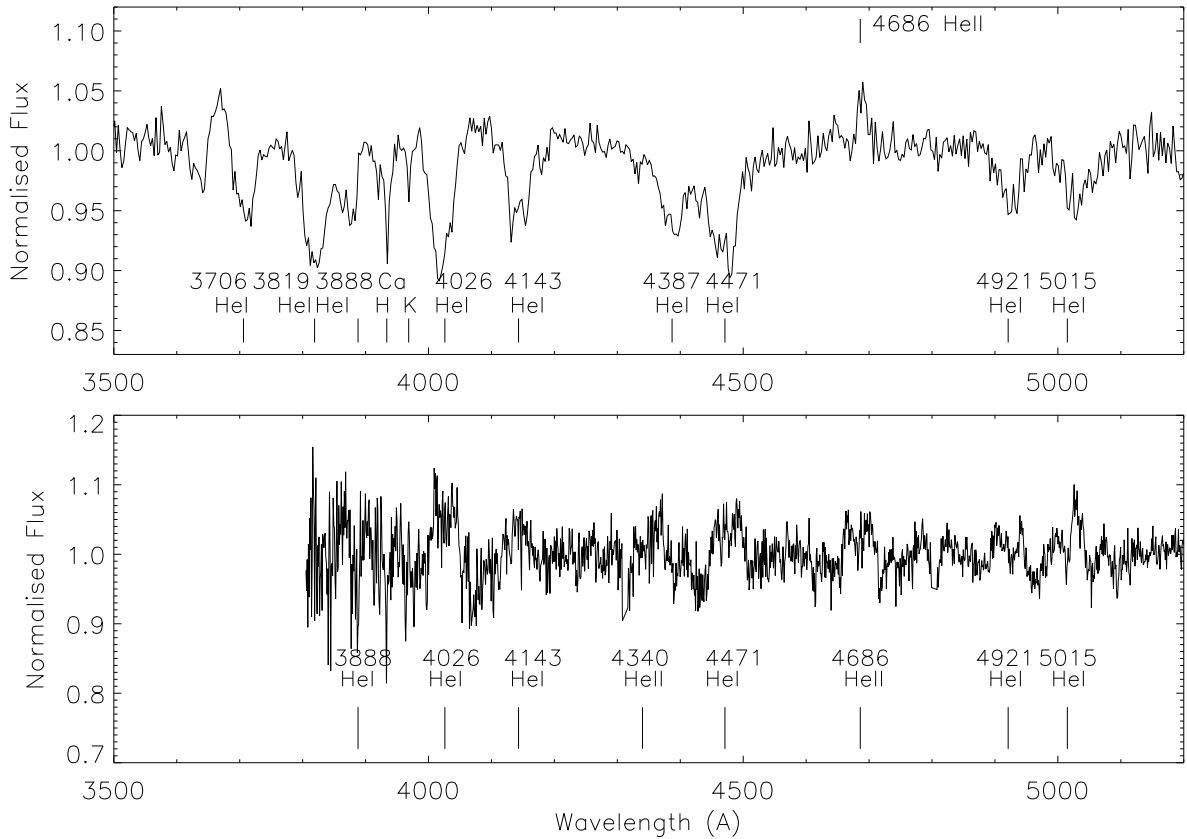}}
\end{picture}
\end{center}
\caption{Top panel: the mean optical spectrum of KL Dra obtained using the 
WHT and the ISIS blue arm when it was in outburst (2009 Oct 10, 
MJD=55114). Lower Panel: the mean optical spectrum of KL Dra obtained
using the Gemini North Telescope when it was in a low state (2007 July
17).}
\label{spec}
\end{figure*}

\subsection{High state spectra}

We were fortunate to be observing on the 4.2m William Herschel
Telescope on La Palma when KL Dra was in outburst. We obtained a
series of 100 sec exposures using ISIS on the night of 2009 Oct 10
which lasted 67 mins.  We used the R300B (giving wavelength coverage
between $\sim$3500--5300\AA) and R158R ($\sim$5200--10000\AA) gratings
together with a 0.8 arcsec slit.  Based on the FWHM of the arc lines,
the resolution of the spectra was determined to be $\sim$3 and $\sim$6
\AA\hspace{2mm} for the blue and red spectra respectively.

We reduced these data using optimal extraction (Horne 1986) as
implemented in the {\sc Pamela}\footnote{ PAMELA was written by T.
Marsh and can be found at http://www.warwick.ac.uk/go/trmarsh} code
(Marsh 1989) which also uses the {\sc Starlink}\footnote{The Starlink
Software Group homepage can be found at
http://starlink.jach.hawaii.edu/starlink} packages {\sc Figaro} and
{\sc Kappa}. The wavelength calibration was linearly interpolated from
copper-argon arc lamp exposures bracketing our observations.

We show the mean spectrum obtained in the blue arm in the upper panel
of Figure \ref{spec}.  Broad absorption lines of He I
(FWHM$\sim$30--35\AA, corresponding to velocities of $\sim$2000 km/s)
are detected, which is similar to the discovery spectrum of KL Dra
(when it was in an outburst) shown in Wood et al. (2002). Unlike the
discovery spectrum we detected an emission line at 4686
\AA\hspace{1mm} (He II). The red spectra are remarkably featureless
with only a possible detection of the He I 5876 \AA\hspace{1mm} and He
II 7177 \AA\hspace{1mm} lines in absorption. We searched for radial
velocity shifts in the absorption lines but found none, with an upper
limit of $\sim$40 km/s. This is consistent with a disc origin for the
lines and a mass ratio typical for AM CVn systems. We also searched
for variability in the flux of the 4686 \AA\hspace{1mm}
line. However, the low signal-to-noise did not allow us to place any
significant constraints on this question.

\subsection{Low state spectra}

We observed KL Dra on the night of 17 July 2007 with the GMOS
instrument in long-slit spectroscopy mode on the Gemini North
Telescope at Mauna Kea. The acquisition image allowed us to determine
its brightness ($g$=19.0) using the local comparison stars mentioned
in \S \ref{phot} and showed it was in a low optical state.

We obtained 60 consecutive 180 sec exposures and used a $1.0$'' slit
and the B1200 grating with binning factors of $4$ (spatial) and $2$
(spectral), giving a wavelength range of $3800-5300$\AA \ with a
dispersion of $0.46$\AA \ per binned pixel.  Weather conditions for
these observations were good, with photometric transparency and
variable seeing around $1^{''}$.

The low-state spectra were reduced in the same manner as the high-state 
spectra. We combine the results into a single averaged spectrum
which we plot in the lower panel of Figure \ref{spec}. In contrast to
the spectrum taken in the high optical state, the low-state spectrum
is dominated by weak emission lines and are typically double-peaked
and indicative of emission from optically-thin regions in the
helium-dominated accretion disc. Moreover, the widths of the lines are
very broad (reaching +/- 1800 km/s in the case of the He 4686 \AA
\hspace{1mm}line) with the presence of an enhancement at velocities
close to zero km/s. This could be connected to the central `spike'
that is seen more prominently in the AM CVn system GP Com which is
thought to originate close to the accreting white dwarf (Marsh 1999).

\section{Swift Observations}

We obtained target of opportunity observations of KL Dra using the
NASA {\swift} satellite (Gehrels et al. 2004) once every two days for
two months. The goal of these observations was to characterise the UV
and X-ray flux of KL Dra over the course of an outburst cycle.
Because {\swift} is in a low Earth orbit, each observation sequence
(which makes up an `ObservationID') is made up of typically 2--4
separate pointings. The exposure time of each UVOT image is generally
600 sec in duration, while the total exposure time of the X-ray
observation in each ObservationID is typically 2--3 ksec. Observations
commenced on 2009 Nov 08 (MJD=55146).

\subsection{UVOT observations}

The Ultra-Violet/Optical Telescope (UVOT) on board the {\swift}
satellite has a 30 cm primary mirror and 6 optical/UV filters (Roming
et al. 2005).  The main goal of our UVOT observations was to determine
how the UV flux changed over the course of an outburst. Since {\swift}
operates a `filter of the day', we were not able to pre-define the
filter, but images were obtained in either the U (central wavelength
347 nm, and a full width half maximum of 79 nm), UVW1 (251 nm, 70 nm),
UVM2 filter (225 nm, 50 nm), or the UVW2 (188nm, 76 nm) filters.

To determine the UV flux of KL Dra we used the {\swift} tool {\tt
uvotmaghist} which is part of the
HEASoft\footnote{http://heasarc.gsfc.nasa.gov/docs/software/lheasoft/}
package of software. This tool takes into account effects such as
coincidence loss and converts the count rate to flux based on
observations of white dwarfs made as part of the {\swift} calibration
process (Poole et al. 2008). As recommended in the UVOT software
guide\footnote{http://swift.gsfc.nasa.gov/docs/swift/analysis/
\newline UVOT\_swguide\_v2.pdf} we used a source radius of
3$^{''}$. We chose a background region free from stars and galaxies.
We note that while the nearby galaxy is much brighter at optical bands
compared to KL Dra, in the UVW1 filter KL Dra is much brighter than
the galaxy and in the UVW2 filter the galaxy is very faint.

We found that in the low optical state the UV flux was dependent on
the UV filter which was used -- eg the flux in the UVW2 filter was
consistently higher than the other UV filters. This is due to the fact
that the filters are sampling different parts of the spectrum of KL
Dra. To put the UV flux on a consistent scale (the implied flux in the
UVW1 filter) we convolved the effective area curves for the different
UVOT filters with white dwarf atmosphere models of different
temperatures (Koester, private communication). For a white dwarf with
$T$=16000 K, this implies a correction factor of 0.68 for the UVW2
data, 1.82 for the U data and 1.23 for the UVM2 data (the results were
were similar when we used blackbodies). The correction factors for
individual filters are comparable within a temperature range of
$\pm$4000 K. If the temperature of the UV component changes
significantly in the outburst then the fluxes will be more uncertain.

We show the UV light curve in the middle panel of Figure
\ref{swift-light}.  The corrected fluxes in the different UVOT filters
now give consistent fluxes during the low optical state.  The first
point to note is that the optical outburst seen at MJD 55173 (2009 Dec
08) is also seen at UV wavelengths and the increase in flux is very
similar (a factor of $\sim$20). Whilst this result was expected, it is
the first time (to our knowledge) that an outburst of an AM CVn system
has been seen in both optical and UV wavebands. It gives confidence
that an outburst seen at UV wavelengths would also be seen at optical
wavelengths.  The data is consistent with the UV and optical flux
starting to increase at the same time (to within half a day). In
contrast, in the case of SS Cyg (which with an orbital period of 6.6
hrs will have a much larger disc than KL Dra) one outburst was
observed to be delayed by 1.5--2.0 days at extreme UV energies
compared to the optical (Wheatley, Mauche \& Mattei 2003).

We also note that the initial rise to maximum brightness is relatively
short and there is a significant decrease in the UV flux (which is
similar to the drop in the optical flux seen at MJD$\sim$55118, Figure
1), after which there is an increase in the UV flux. We note a brief UV
brightening seen at MJD$\sim$55159 which was also marked by a small
increase in the optical brightness.

Given that our ground-based optical observations gave some indication
that the outbursts repeated on a timescale of $\sim$60 days, we
obtained a further set of {\swift} observations starting on 2010 Feb
09. As expected, KL Dra was in a bright UV state (cf, Figure
\ref{swift-light}). The data folded on a period of 61.5 days (Figure
\ref{fold-long}) shows a very similar profile to that of the optical
data.

\subsection{XRT observations}

The X-ray Telescope (XRT) (Burrows et al. 2005) on-board {\swift} has
a field of view of 23.6$\times$23.6 arcmin with CCD detectors
allowing spectral information of X-ray sources to be determined. It is
sensitive over the range 0.2-10keV and has an effective area of
$\sim$70 cm$^{2}$ at 1keV (for comparison the {\sl Rosat} XRT had an
effective area of $\sim$400 cm$^{2}$ at 1keV).

The XRT has a number of modes of operation (designed to observe
gamma-ray bursts at various stages of their evolution) but here we
concentrate on the `photon counting' mode which has full imaging and
spectroscopic information. The data are processed using the standard
XRT pipeline and it is these higher level products which we use in our
analysis {\footnote{http://swift.gsfc.nasa.gov/docs/swift/analysis/
\newline xrt\_swguide\_v1\_2.pdf}.

To determine the count rate of KL Dra at each epoch we combined the
X-ray events from each ObservationID into a corresponding image using
{\tt
xselect}\footnote{http://heasarc.nasa.gov/docs/software/lheasoft/ftools/
\newline xselect/xselect.html}. This image was input to the HEASoft
tool {\tt XIMAGE} and the routine {\tt SOSTA} (which takes into
account effects such as vignetting, exposure and the point spread
function) to determine the count rate and error at the position of KL
Dra. We show the count rates for the individual observations in the
lower panel of Figure \ref{swift-light}. Since the count rates are
low, we created images using more than one observation and determined
the count rate from these. We also show these results in the lower
panel of Figure \ref{swift-light} as thicker symbols.

In the lower panel of Figure \ref{fold-long} we show the X-ray data
folded on a 61.5 day period. We created an image from all the X-ray
data taken in the low optical state and also the high optical
state. The mean count rates (0.00258$\pm$0.00029 ct/s for the low
optical state and 0.00190$\pm$0.00036 ct/s for the high optical state)
are over-plotted as thicker lines in Figure \ref{fold-long}. Although
the mean X-ray flux during the low optical state is higher than that
of the high optical state the difference is not significant: we find
no evidence that the X-ray flux changed significantly between the high
and low optical states. We also tested whether there was a change in
the soft/hard (0.1--1keV/1--10keV) ratio between the low and high
optical states -- there was none.

\begin{figure}
\begin{center}
\setlength{\unitlength}{1cm}
\begin{picture}(8,12.5)
\put(-0.8,-0.8){\includegraphics{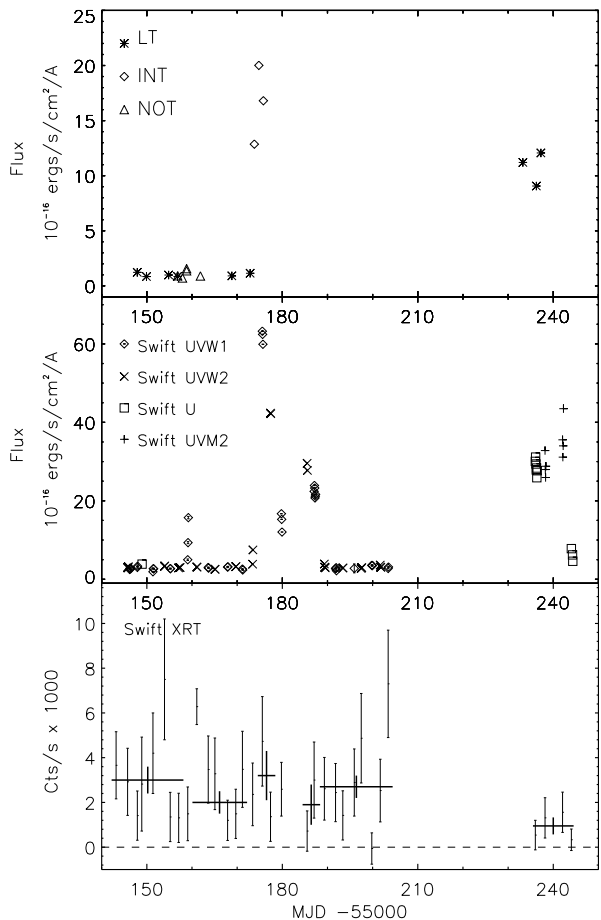}}
\end{picture}
\end{center}
\caption{The light curves of KL Dra made over the course of the 
{\swift} observations. Top panel: Ground-based $g$ band observations made 
using the LT, INT and NOT. In order to make a suitable comparison with 
the UV and X-ray data we converted the $g$ band mag to flux using the 
conversion quoted in Holberg \& Bergeron (2006). 
Middle Panel: The UV flux (in units of 10$^{-16}$ {\ergscm} \AA$^{-1}$) where
we have normalised the flux in the different filters to match that expected
in the UVW1 filter (see text for details).
Bottom Panel: The X-ray count rates as determined using the {\swift} XRT.}
\label{swift-light}
\end{figure}

\begin{figure}
\begin{center}
\setlength{\unitlength}{1cm}
\begin{picture}(8,12.5)
\put(-0.8,-0.8){\includegraphics{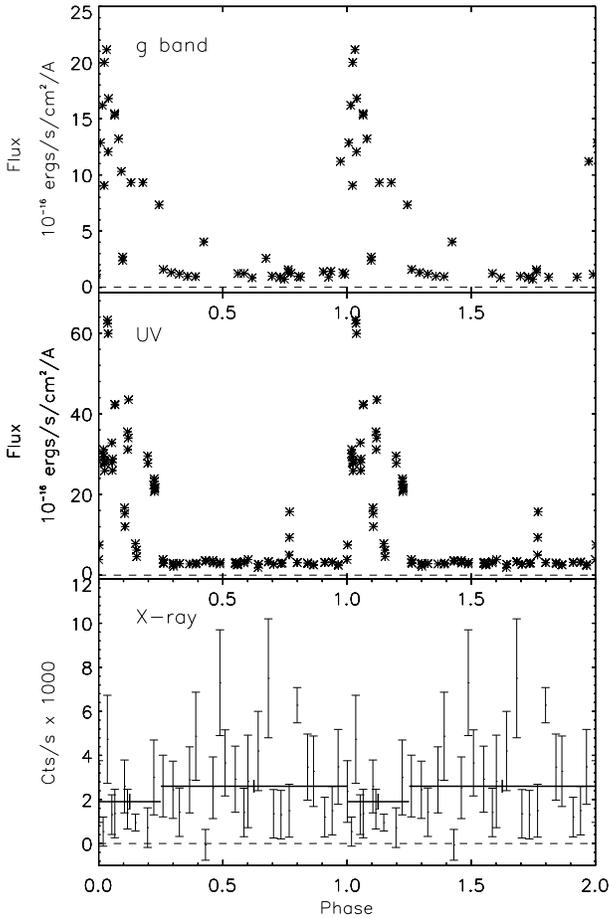}}
\end{picture}
\end{center}
\caption{The light curves of KL Dra made at three wavebands folded on a period
of 61.5 days with a $T_{o}$= MJD 55173.4.}
\label{fold-long}
\end{figure}

Using {\tt XSELECT} we initially extracted an X-ray spectrum of KL Dra
over a time interval when it was in a low optical state. In addition,
we extracted a background spectrum from a source free region. We used
the appropriate response matrix from the {\swift} calibration files
and created an auxiliary file using the HEASoft tool {\tt xrtmkarf}.

We fitted the X-ray spectrum of KL Dra in a low optical state using
the {\tt vmekal} thermal plasma model and the {\tt tbabs} neutral
absorption model. Keeping the metallicity fixed at solar (but with the
hydrogen abundance fixed at zero) we found a good fit to the data
(\rchi=0.84, 6 degrees of freedom). The hydrogen column density
determined from our model fits is consistent with the total hydrogen
column density to the edge of the Galaxy in the direction of KL Dra
($\sim7.4\times10^{20}$ \pcmsq. Dickey \& Lockman 1990).  Since, we
found no evidence that the X-ray flux varied between the low and high
optical states we created a second spectrum using all the X-ray data.
Using a model in which the metallicity was fixed at solar we obtained
a fit with \rchi=1.12, 8 dof. We give the fits with associated errors
in Table \ref{xray-fits}.

\begin{table}
\begin{center}
\begin{tabular}{lr}
\hline
$N_{\rm H}$ & $0.05^{+0.11}_{-0.05}\times10^{20}$ (\pcmsq) \\
$kT$ & 2.8$^{+4.1}_{-1.0}$ keV \\
Flux$_{\rm o}$ & 7.7$^{+2.1}_{-1.8}\times10^{-14}$ (\ergscm) (0.1-10keV)\\
Flux$_{\rm u}$ & 10.4$^{+2.8}_{-2.5}\times10^{-14}$ (\ergscm) (0.01-100keV)\\
& {\rchi}=1.12 (8 dof) \\
\hline
\end{tabular}
\end{center}
\caption{The spectral fits to the X-ray spectrum of KL Dra. We used the 
{\tt tbabs} absorption model and the {\tt vmekal} thermal plasma model in 
{\tt XSPEC}. Flux$_{\rm o}$ refers to the observed flux while Flux$_{\rm u}$
refers to the unabsorbed flux.}
\label{xray-fits}
\end{table}

\section{Distance and Luminosity}
\label{distance}

Ramsay et al. (2006) determined the X-ray and UV luminosities for 8 AM
CVn systems: of those, 5 had distances determined using parallax
measurements.  The X-ray luminosity of those 5 systems decreases as
the orbital period increases. Based on this trend, we predict that a
system with an orbital period of 25 mins should have an X-ray
luminosity of $L_{\rm X}\sim5\times10^{30}$ \ergss. This is very
similar to CR Boo (5.2$\times10^{30}$ \ergss, Ramsay et al. 2006) with
has an orbital period very close to KL Dra. Our findings imply that KL
Dra lies at a distance of 550--850pc (using the standard error on the
unabsorbed bolometric X-ray flux, Table \ref{xray-fits}).

The UV luminosity is more uncertain since it is difficult to constrain
the temperature of the UV component (which derives from the primary
white dwarf plus the accretion disc). However, taking the lead from
Ramsay et al. (2006), we fix a single blackbody with a range of
temperature (10000--40000K), and fix the normalisation of the
blackbody so that it matches the measured UV flux (we used the {\tt
uvred} absorption component in {\tt XSPEC}, Arnaud 1996).  Setting the
UV flux near the UV maximum ($6\times10^{-15}$ \ergscm, Figure
\ref{swift-light}) we find a UV luminosity of 3--10$\times10^{33}$
\ergss for a distance of 700pc and a temperature range 10000--40000K.
For a low optical state the UV luminosity reduces by a factor of 20 to
$\sim1-5\times10^{32}$ \ergss, which is still greater than the X-ray
luminosity by a factor of 20.

During the low state, the UV flux originates from the accreting white
dwarf and the accretion disk. It is therefore difficult to determine
the mass accretion rate (which is less than the mass transfer rate,
cf. Schoembs \& Hartmann 1983) during the low state.  However, if we
use the high state UV luminosity and $L=GM_{1}\dot{M}_{\rm
acc}/R_{1}$, we obtain $\dot{M}_{\rm acc}\sim3-10\times10^{16}$ g/s
(=$4.7-15.7\times10^{-10}$ \Msun/yr, assuming $M_{1}$=0.6 \Msun, and a
distance of 550-850 pc). Deloye et al. (2007) predict the mass
accretion rate as a function of the mass and structure of the mass
donor star and whether it is irradiated and find $\dot{M}_{\rm
acc}\sim5-20\times10^{15}$ g/s (which is slightly lower compared with
that determined using the high state UV luminosity) for an orbital
period of 25 mins. A greater understanding of the origin of the UV
emission during the outburst cycle and the nature of the mass donor
star is required before the predicted mass accretion rate can be
usefully compared with observations.

\section{Modelling the light curves}

We model our optical lightcurve of KL Dra in the framework of the Disc
Instability Model (hereafter DIM, e.g. Hameury et al. 1998) which has
been adapted for helium discs by Lasota, Dubus \& Kruk (2008). (See
Lasota 2001 for a review of the DIM). We calculated model lightcurves
for several different sets of parameters: $\alpha_{\rm C}$ (the
viscosity parameter of the disc in the cold state), $\alpha_{\rm H}$
(the hot state), and the mass transfer rate, $\dot{M}_{\rm tr}$. In
our simulations we fixed the mass of the accreting white dwarf at 0.6
\Msun. We also allowed the inner radius of the accretion disc, $r_{\rm
in}$, to approach the radius of the white dwarf. The outer radius of
the disc, $r_{\rm out}$, was allowed to vary around a mean of $r_{\rm
out}=1.2\times10^{10}\,\mathrm{cm}$.

For our initial simulations we took $\dot{M}_{\rm tr}=3\times10^{16}$
g/s and we show a simulated light curve in Figure \ref{dim} using
$\alpha_{\rm C}$=0.025 and $\alpha_{\rm H}$=0.026.  It shows outbursts
which repeat on a $\sim$60 day timescale, an amplitude of $\sim$3 mag
and bright state lasting $\sim$2 weeks. In the simulations there is an
increase in flux of $\sim$0.5 mag between the end of one outburst and
the start of the next: this is a known deficiency of the DIM (see
comments in Lasota 2001). In contrast, if we assume $\dot{M}_{\rm
tr}=1\times10^{17}$ g/s in our simulations then this low-state flux
increase is much greater. Further, they give more asymmetric outburst
profiles (with the decline being more extended).

We also simulated a set of light curves where we assumed that the 
secondary star was irradiated by X-rays
and UV photons emitted in the accretion region and from the white
dwarf. Irradiation of the secondary can cause an enhancement of the
mass transfer rate. We found that although this gave values for the 
viscosity parameters similar to that of hydrogen-dominated accreting 
dwarf novae (eg $\alpha_{\rm C}\sim$0.02 and $\alpha_{\rm H}\sim$0.1--0.2,
Hameury et al. 1998, Smak 1999) the simulated light curves gave a large 
increase in brightness (over 1 mag) between the end of an outburst and 
the next. A more detailed investigation can
help elucidate how different conditions (different critical values of
$\Sigma$ and $T$, smaller disc sizes) could affect cold and hot
viscosity parameters.

\begin{figure}
\begin{center}
\setlength{\unitlength}{1cm}
\begin{picture}(8,7)
\put(-1.5,-6.3){\includegraphics{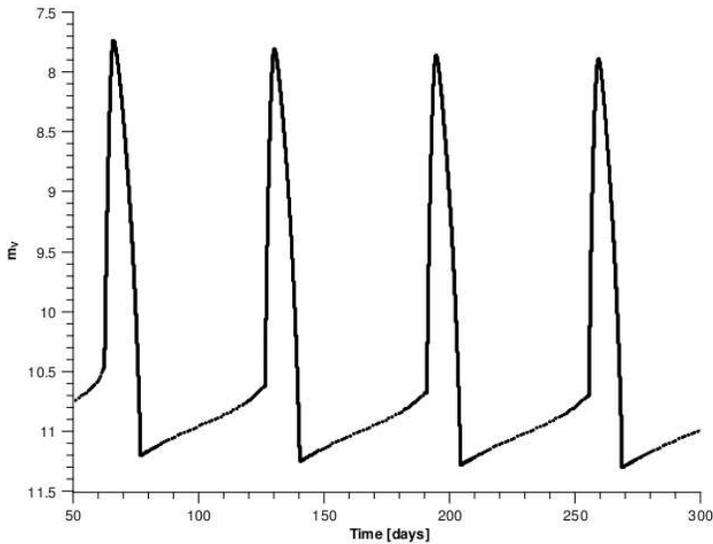}}
\end{picture}
\end{center}
\caption{The simulated light curve using the Disc Instability Model which
uses a helium accretion flow.}
\label{dim}
\end{figure}

\section{Discussion}

\subsection{KL Dra as a helium dwarf nova}

The AM CVn systems CR Boo and V803 Cen have been known for several
decades and hence their optical light curves have been well studied --
they also have orbital periods close to that of KL Dra. The light
curves of all three systems have shown a photometric signal on the
`superhump' period (a signature of the precession period of the
accretion disc) when the system is in a high optical state and is a
few percent longer than the orbital period.

On longer timescales, both V803 Cen (Patterson et al. 2000) and CR Boo
(Patterson et al. 1997) have shown a prominent modulation ($\sim$1 mag)
on a period of $\sim$19--22 hrs at various epochs: this has been
called the `cycling state'.  Patterson et al. (2000) noted that the
amplitude and period of these modulations placed these systems on the
short period end of the Kukarkin-Parenago period-amplitude
relationship of normal dwarf novae outbursts (Warner 1987). They
concluded that these near day-long modulations were similar to
hydrogen-rich dwarf novae outbursts (in particular the short-period ER
UMa systems).  On even longer timescales, evidence has been found for
a period of 77 days in V803 Cen (Kato et al. 2000a) and a period of 46
days in CR Boo (Kato et al. 2000b). Both binaries spend around half the
time in a bright state.

Whilst our observations of KL Dra are less extensive than those of
V803 Cen and CR Boo, they show characteristics which are clearly
different to these two AM CVn systems. Indeed, with a possible
outburst cycle of $\sim$60 days and with a duty-cycle of $\sim$1/5, KL
Dra closely resembles the outbursts seen in the SU UMa or U Gem type
dwarf novae. However, at this stage we cannot exclude that the
outbursts seen in KL Dra are not the AM CVn equivalent of WZ Sge
outbursts (dwarf novae which recur on a timescales of years and have
outburst amplitudes of 6--9 mag, eg Patterson et al. 2002).  As such
it makes KL Dra a prime target to investigate the similarities and
differences between hydrogen-rich and hydrogen-deficient accretion
flows.

\subsection{The X-ray flux over the outburst cycle} 

The hydrogen-accreting dwarf novae have been much studied at various
wavelengths over several decades and show regular outbursts on
timescales ranging from a few weeks to months or even years.  Many
systems show characteristics similar to SS Cyg which showed in
one outburst that the increase in the extreme UV emission was delayed
by 1.5--2.0 days with respect to the optical emission (Wheatley,
Mauche \& Mattei 2003). In contrast, the hard X-rays, after an initial
burst, were suppressed during the remainder of the outburst. To
complicate matters, U Gem has shown at least one outburst where the
hard X-rays follow the optical and UV flux (Mattei, Mauche \& Wheatley
2000).  It is not clear if this is due to the relatively high binary
inclination of U Gem. More recently GW Lib (which has only had two
known outbursts) also shows the enhancement of X-rays at the start of
the outburst (Byckling et al. 2009).

One of the key reasons for obtaining {\swift} observations was to
determine how the X-ray flux of KL Dra varied over the course of the
outburst cycle. It is clear that X-rays are detected from KL Dra both
during the low and the high optical states. However, there is no
evidence that the X-ray flux changed significantly during the course
of the outburst (Figure \ref{swift-light}).

\section{Conclusions}

Our observations of KL Dra show that it undergoes frequent optical
outbursts which are also seen at UV wavelengths. Although our coverage
is by no means complete, our data are consistent with KL Dra
undergoing outbursts on a period of $\sim$60 days. The amplitude of
the outbursts ($\sim$3 mag) and duration of the outburst ($\sim$2
weeks) are very similar to the typical outbursts that seen in the SU
UMa or U Gem sub-class of (hydrogen) accreting dwarf novae.  As such
we encourage other observers, whether they are amateur astronomers who
have suitable equipment or team members of survey telescopes which
have suitable spatial resolution (to resolve KL Dra from its `nearby'
galaxy), to obtain a more complete long term coverage and hence verify
the $\sim$60 day outburst period.  The fact that KL Dra shows such
regular outbursts makes it an ideal system with which to investigate
helium accretion flows in detail.

\section{Acknowledgements}

The Liverpool Telescope is operated on the island of La Palma by
Liverpool John Moores University with financial support from the UK
Science and Technology Facilities Council. We thank the LT Support
Astronomer Chris Moss for his assistance in scheduling our
observations and the {\swift} PI, Neil Gehrels, together with the
{\swift} science and operations teams for their support of these
observations.  We also thank Pasi Hakala and Tiina Liimets for
obtaining several images using the Nordic Optical Telescope which is
operated on the island of La Palma jointly by Denmark, Finland,
Iceland, Norway, and Sweden. The Isaac Newton and William Herschel
Telescopes are operated on the island of La Palma by the Isaac Newton
Group.  All three telescopes are located in the Spanish Observatorio
del Roque de los Muchachos of the Instituto de Astrofisica de
Canarias.  Observations were also obtained at the Gemini Observatory,
which is operated by the Association of Universities for Research in
Astronomy, Inc., under a cooperative agreement with the NSF on behalf
of the Gemini partnership: the National Science Foundation (United
States), the Science and Technology Facilities Council (United
Kingdom), the National Research Council (Canada), CONICYT (Chile), the
Australian Research Council (Australia), Ministério da Ciência e
Tecnologia (Brazil) and Ministerio de Ciencia, Tecnología e Innovación
Productiva (Argentina).  We thank Jean Pierre Lasota for useful
comments on a previous draft of this paper. DS acknowledges a STFC
Advanced Fellowship.

\end{document}